# Архивы спектральных наблюдений КрАО. Каталоги объектов и изображений

*Шляпников А.А.[1], Горбунов М.А.[1], Горбачев М.А.[1,2], Акметдинов Р.Р.[3]*

[1] ФГБУН «Крымская астрофизическая обсерватория РАН», Научный, Крым, Россия, 98409
*aas@craocrimea.ru, mag@craocrimea.ru*
[2] ФГАОУВО «Казанский федеральный университет», Кремлёвская, 18, Казань, Россия, 420008
*mark-gorbachev@rambler.ru*
[3] ФГБОУВО «МГУ имени М.В. Ломоносова», Ленинские горы, 1, Москва, Россия, 119991
*rusak812@gmail.com*

**Аннотация.** Работа, описанная в данной статье, является продолжением ранее начатых исследований по архивным спектральным наблюдениям, выполненным в Крыму. Она охватывает интервал времени около 90 лет и содержит информацию о спектроскопии с использованием: от широкоугольных астрографов с объективной призмой до главного телескопа КрАО - ЗТШ. Рассмотрена краткая история инструментов, их оборудование. Статья проиллюстрирована возможностями сетевого доступа к каталогам наблюдений на различных инструментах в интерактивном атласе неба Aladin с переадресацией к оригинальным спектрограммам. Для них выполнено преобразование линейных координат отсканированных негативов в шкалу, соответствующую длинам волн. Показаны возможности учёта спектральной чувствительности регистрируемых изображений по абсолютному распределению энергии. Особенностью представляемой работы является связь оцифрованных оригинальных наблюдений и результатов их независимой обработки с данными, опубликованными для объектов в "Известиях Крымской астрофизической обсерватории".

ARCHIVES OF CrAO SPECTRAL OBSERVATIONS. CATALOGUES OF OBJECTS AND IMAGES, *by A.A. Shlyapnikov, R.R. Akmetdinov, M.A. Gorbunov and M.A. Gorbachev.* The work described in this article is a continuation of the previously initiated research on archival spectral observations carried out in the Crimea. It covers a time interval of about 90 years and contains information about spectroscopy using various facilities: from the wide-angle astrographs with an objective prism to the main CrAO telescope - ZTSh. A brief history of telescopes and their equipment are presented. The article is illustrated with the possibilities of network access to the catalogues of observations taken with various instruments in the interactive Aladin Sky Atlas with the redirection to the original spectrograms. For this aim, the linear coordinates of the scanned negatives were converted into a scale which corresponds to the wavelengths. The possibilities of taking into account the spectral sensitivity of the recorded images by the absolute energy distribution are shown. A feature of this work is the connection of digitized original observations and the results of their independent processing with the data published for objects in the "Izvestiya of the Crimean Astrophysical Observatory" journals.

**Ключевые слова:** архивы наблюдений, базы данных, спектры, каталоги

## 1 Введение

Обширный архив спектральных наблюдений накоплен за более чем вековую историю астрономических наблюдений в Крымской астрофизической обсерватории (КрАО). Он включает в себя, как фотопластинки и фотоплёнки, полученные на астрографах с объективной призмой и на телескопах с различными спектрографами, так и фотоэлектрические, телевизионные и ПЗС записи (Полосухина и др. 1998; Горбунов, Шляпников 2013; Пакуляк и др. 2014).

Для большей части архивов составлено описание и подготовлены списки наблюдений и каталоги объектов в цифровом формате. Некоторые из них размещены на сервере КрАО и доступны в сетевом режиме. В данной работе собраны некоторые рекомендации по использованию спектральных архивов обсерватории.

## 2 Краткое описание архивов
### 2.1. Спектроскопия с объективной призмой[1]

Коллекция спектральных наблюдений, выполненных с объективной призмой, состоит из трех частей, что обусловлено инструментами, на которых были получены негативы. В таблице 1 приведены основные характеристики астрографов и дисперсия для оцифрованных спектрограмм.

Таблица 1

| Название спектрографа | Диаметр объектива (мм) | Фокусное расстояние (мм) | Дисперсия у линии $H_\gamma$ (Å/pix) |
|---|---|---|---|
| **Унар (Unar)** | 117 | 600 | 1.5 |
| **Догмар (Dogmar)** | 167 | 750 | 1.4 |
| **400-mm** | 400 | 1600 | 1.5 |

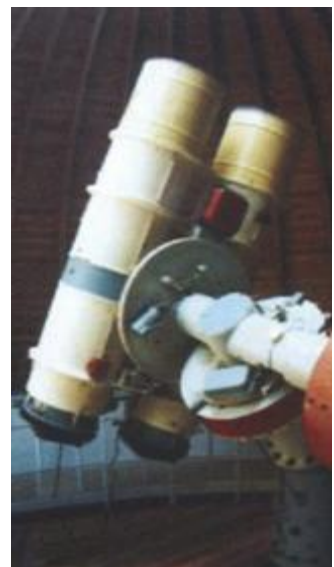

*СОПКА* содержит около 1000 спектральных пластин, полученных на астрографах с объективами «Унар», «Догмар» и 400-мм (рис. 1) в период с 1929 по 1965 годы. Большую часть коллекции составляют негативы с наблюдениями, выполненными по «Плану академика Г.А. Шайна»[2], организованному для изучения структуры Галактики в 1950-1965 годах (Проник, 2005). По результатам реализации этого проекта было составлено 13 каталогов, содержащих около 35 тысяч звезд, с определёнными фотометрическими и спектральными характеристиками (Горбунов, Шляпников, 2017а; Горбунов, Шляпников, 2017б).

Рис. 1. Двойной 400-мм астрограф КрАО, использованный при реализации «Плана академика Г.А. Шайна». В передней части одного из телескопов укреплялась объективная призма, в результате чего в фокальной плоскости астрографа получались изображения спектров звёзд (см. рис. 3). Второй астрограф применялся для регистрации прямых изображений. Комбинация эмульсий позволяла проводить фотометрию в трёх полосах.

Рис. 1.

---

[1] Сокращения, используемы в ресурсе Крымской Астрономической Виртуальной Обсерватории (КрАВО): «Спектроскопия с Объективной Призмой. Коллекция Астронегативов» (КрАВО *СОПКА*), или английский вариант «Spectroscopy with Objective Prism» (CrAVO *SwOP*).
[2] «План академика Г.А. Шайна» на сервере КрАО: http://www.crao.ru/~aas/PROJECTs/SPPOSS/SPPOSS.html

Рисунок 2 иллюстрирует покрытие небесной сферы фотографическими пластинами из коллекции *СОПКА*. В основном, негативы получены вдоль Млечного пути.

На рисунке 3 приведен фрагмент негатива из архива *СОПКА*, снятого с объективной призмой, и загруженного в интерактивный атлас неба Aladin (Боннарель и др., 2000). Выделены спектры, описанные в разделе 3.2.

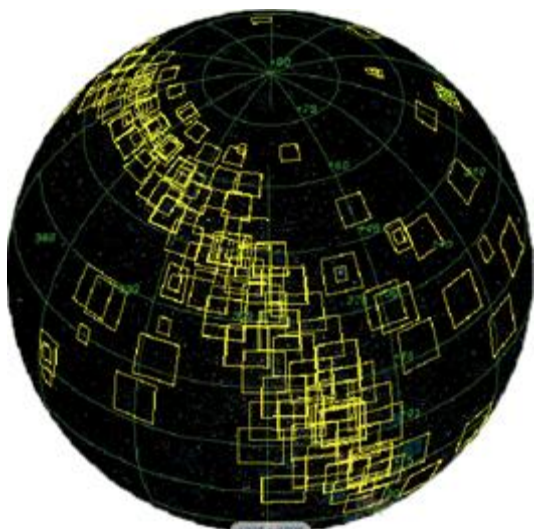

Рис. 2. Распределение *СОПКА* на небесной сфере.

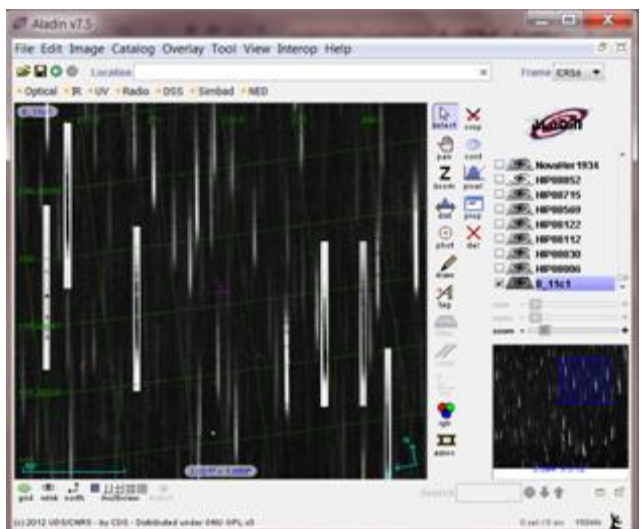

Рис. 3. Фрагмент полноформатного изображения (позитив), с выделенными объектами (негатив) для последующей обработки.

## 2.2. Спектроскопия с щелевыми спектрографами. Краткое описание
### 2.2.1. 40-дюймовый рефлектор фирмы "Goward Grabb"

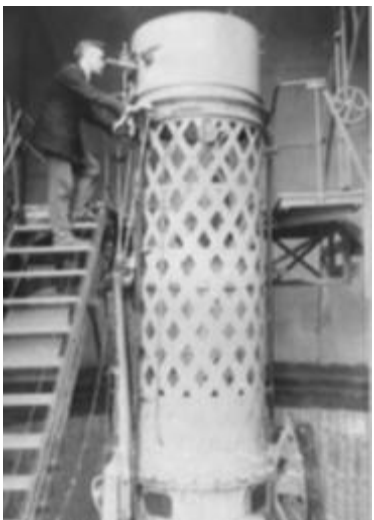

Рис. 4. 40" рефлектор.

40-дюймовый (1000-мм) телескоп (рис. 4) для Симеизского, тогдашнего отделения Главной Пулковской обсерватории, был заказан в 1912 году английской фирме "Goward Grabb". Сначала Первая мировая война, а затем последовавшие за ней события в Российской империи, задержали доставку телескопа в Крым на 13 лет. Монтаж телескопа был начат в октябре 1925 года, а 28 мая 1926 года на нём был получен первый снимок. Телескоп имел фокус Ньютона – 5200-мм и фокус Кассегрена – 18600-мм. В последствие, 40-дюймовый рефлектор был оснащён большим однопризменным спектрографом с термостатом и двумя камерами, имеющими разное фокусное расстояние, кварцевым спектрографом для ньютоновского фокуса и кассетой для прямых снимков (Шайн, 1926; Крючков и др., 2009).

Основной программой исследований на Симеизском 1000-мм рефлекторе стало наблюдение двойных звёзд и определение лучевых скоростей звёзд (Шайн, 1929; Шайн, Альбитский, 1932).

Телескоп был вывезен в Германию во время Второй мировой войны, после Победы обнаружен, но оказался непригодным для восстановления.

### 2.2.2. 48-дюймовый (1220-мм) рефлектор фирмы Цейса (Zeiss-50″)[3]

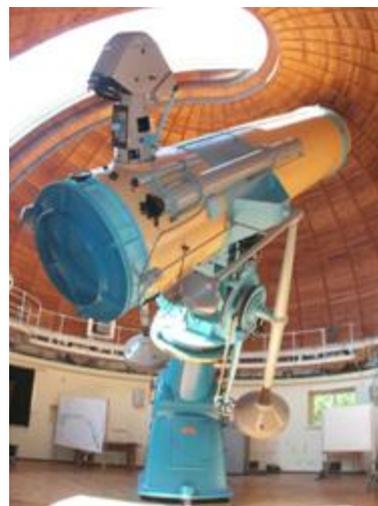

48-дюймовый (1220-мм) телескоп (рис. 5) был изготовлен немецкой фирмой Карл Цейсс для Берлинско-Бабельсбергской обсерватории в начале XX века. Из-за Первой мировой войны и последовавшего за ней мирового кризиса телескоп начал свои регулярные наблюдения только в 1924г. 48-дюймовый рефлектор оставался крупнейшим телескопом Европы в течение 20 лет (с 1924 по 1944г). После Второй мировой войны телескоп был перемещён в Крымскую астрофизическую обсерваторию, взамен разрушенного метрового телескопа Симеизской обсерватории.

Телескоп имеет систему Кассегрена-Нэсмита с эквивалентным фокусным расстоянием 24000 мм. Был снабжён тремя спектрографами: большим «стеклянным», кварцевым и дифракционным АСП-11. Последний спектрограф, работал и в сочетании с электронно-оптическим преобразователем ФКТ-1А.

Детальная информация об исследовании 48″ телескопа приведена в статье И.М. Копылова (1954), а история инструмента описана К.Н. Гранкиным (2013). Астрономические исследования на телескопе, выполнявшиеся в КрАО, представлены в статьях А.А. Боярчука (2013) и Т.М. Рачковской (2013).

Рис. 5. 48″ рефлектор.

Вторая часть 109 тома «Известий КрАО» содержит материалы международной конференции «Телескоп Zeiss-50»: первые сто лет на службе астрономии», которая прошла в обсерватории в 2012 г.

### 2.2.3. 2600-мм телескоп имени академика Г.А. Шайна (ЗТШ)

Телескоп был построен в 1961 году Ленинградским Оптико-Механическим Объединением (ЛОМО) и назван в честь академика Г.А. Шайна. ЗТШ – наибольший оптический телескоп КрАО (рис. 6). Инструмент имеет четыре оптические схемы, в которых было установлено следующее оборудование. Прямой фокус - 9965 мм бесщелевые спектрографы: СП-79, СП-80 и СП-110. Система Кассегрена - 42500 мм дифракционный ПЗС спектрометр. Система Нэсмита - 40750 мм. Спектрографы: СП-72, СПЭМ с ЭОПом и ПЗС. Система куде (прямой фокус) - 104250 мм. Спектрографы: АСП-14, дифракционный ПЗС спектрометр, звездный спектрограф с эшеле.

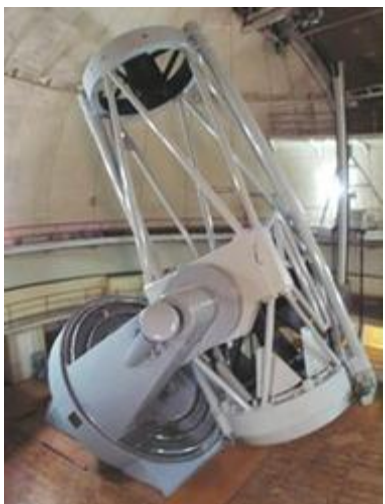

Рис. 6. Телескоп ЗТШ.

Спектральный архив ЗТШ за период с 1964 по 2000 годы содержит информацию о 378 объектах наблюдавшихся фотографическим способом (приблизительно по 3000 записей для прямых снимков и для регистрации с ЭОПом), и о 845 объектах, спектры которых были получены с ПЗС детекторами (около 50000 изображений). На рис. 9 фотографические наблюдения обозначены желтыми маркерами, а ПЗС – красными окружностями.

---

[3] 1220-мм рефлектор в КрАО часто называли «Пятидесяткой», округляя его 48 дюймов до 50.

## 2.2.4. Архивы наблюдений и базы данных

Спектральные фотографические архивы наблюдений, выполненных на трёх, описанных выше телескопах содержатся в стеклянной библиотеке КрАО. В основе баз данных лежат журналы наблюдений и публикации результатов их обработки.

База данных спектроскопии на Симеизском 1-м телескопе содержит более 3000 записей. На основе неё создан *ajs* (Aladin Java Script) файл для 491 объектов, который обеспечивает доступ к информации о звёздах с помощью программы Aladin (рис. 7).

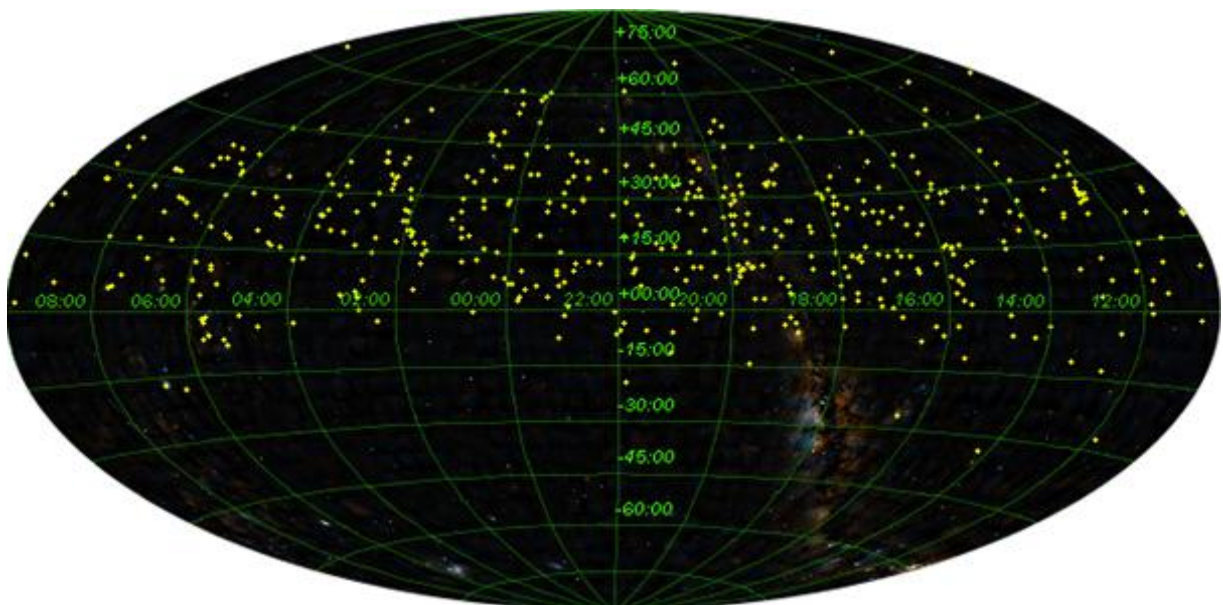

Рис. 7. Распределение на небесной сфере объектов, наблюдавшихся на 1000-мм телескопе.

В 2013 году А.А. Шляпниковым была опубликована статья ««Небо пятидесятки»: каталог и библиография объектов, наблюдавшихся на Zeiss 50". Обновляемая версия». В базе данных спектральных наблюдений 5570 записей. Для доступа в Aladin создан *ajs* файл, в который вошли 808 объектов (рис. 8).

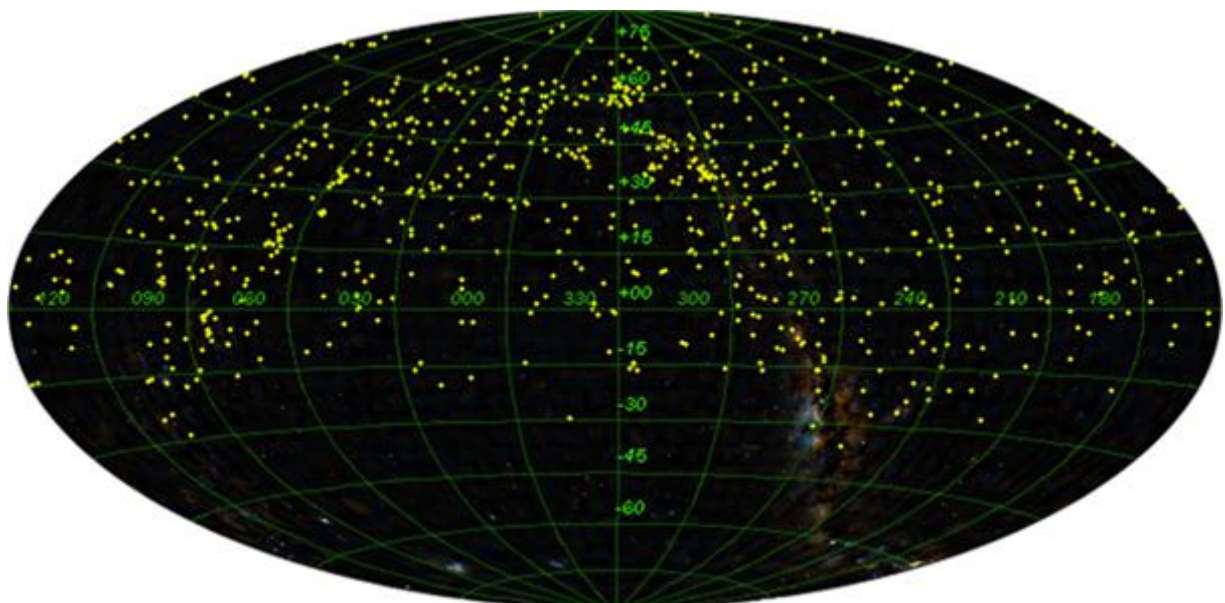

Рис. 8. Распределение на небесной сфере объектов, наблюдавшихся на 48″ телескопе.

База данных спектральных фотографических наблюдений на ЗТШ включает 1223 записи об объектах. Информация доступна в интерактивном атласе неба Aladin (рис. 9).

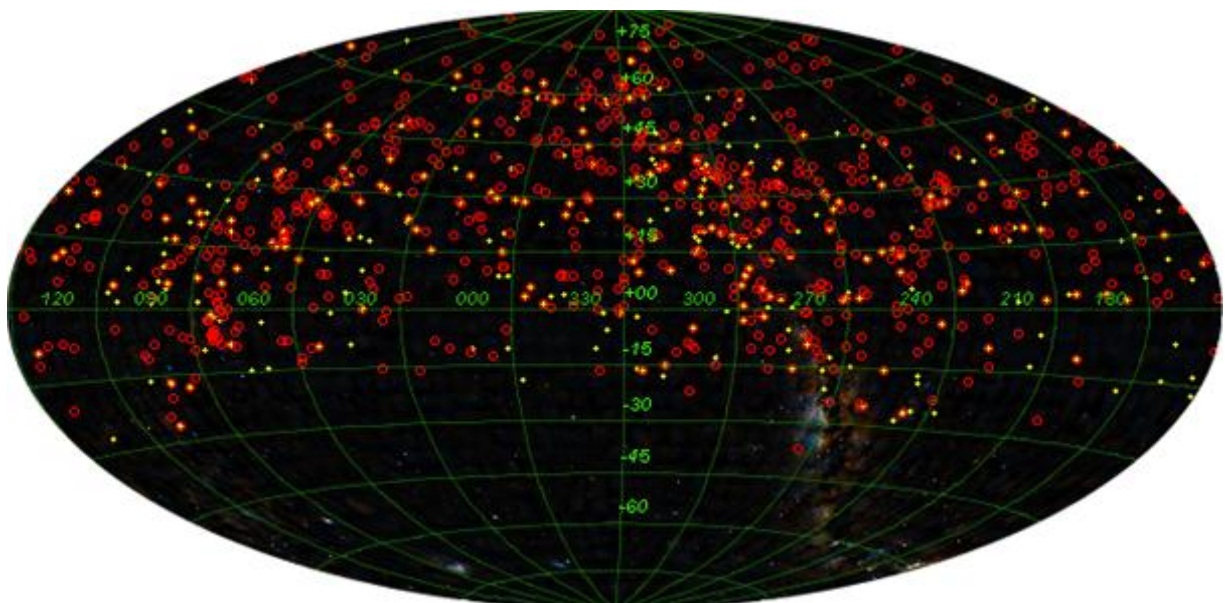

Рис. 9. Распределение на небе объектов ЗТШ (пояснения в тексте).

## 3. Калибровки и примеры цифровых версий спектров
### 3.1. Дисперсионные кривые и спектральная чувствительность

Калибровки по длинам волн цифровых версий наблюдений, выполненных с объективной призмой, производятся по наиболее характерным спектральным линиям. Для щелевых призменных спектрограмм – по спектру сравнения. Примеры таких дисперсионных кривых приведены на рисунках 10 и 11.

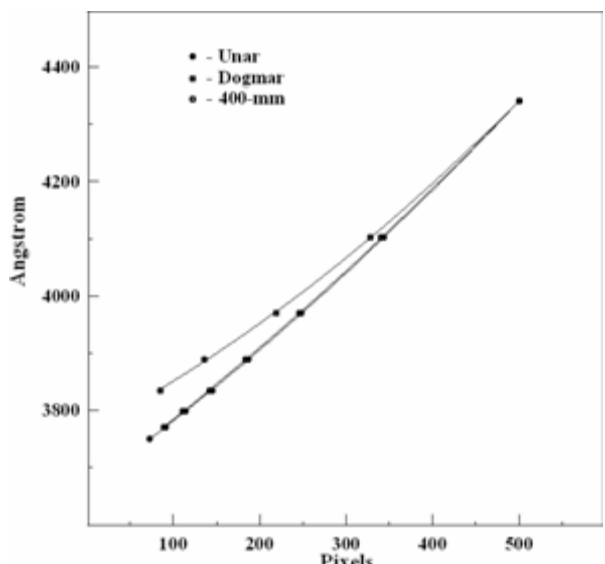
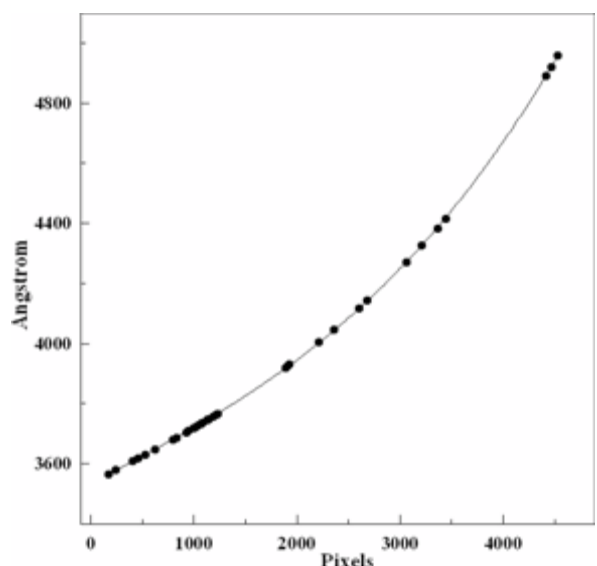

Рис. 10.   Рис. 11.

Рисунок 12 иллюстрирует сравнение нормированных данных спектрофотометрического стандарта (чёрная линия) с данными, извлечёнными для объекта из оцифрованного негатива (красная линия). А рисунок 13 – определенную спектральную чувствительность изображения

А.А. Шляпников, М.А. Горбунов, М.А. Горбачев, Р.Р. Акметдинов

(по данным рис. 12) для дальнейшей редукции спектров, извлечённых из негативов, полученных с объективной призмой.

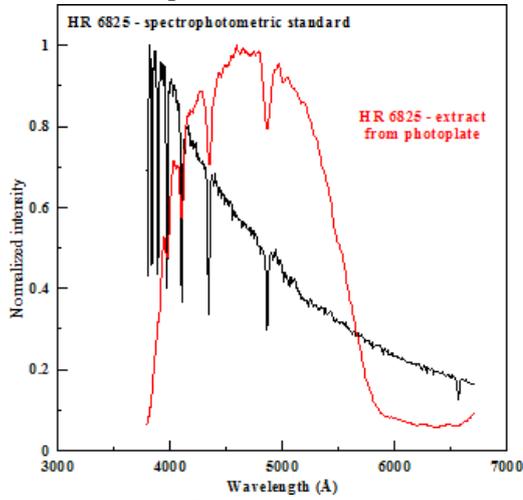
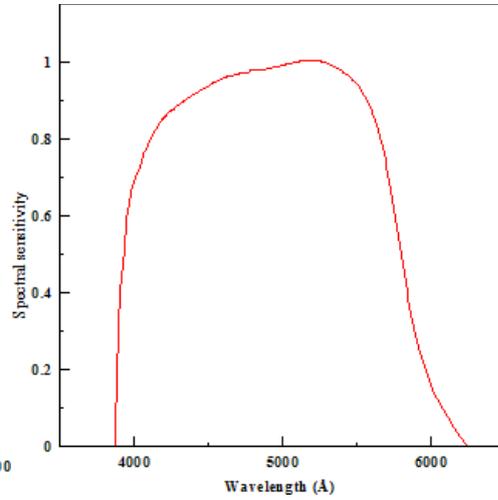

Рис. 12.                                                                                           Рис. 13.

### 3.2. Редукция шкалы изображения спектра низкой дисперсии в ангстремы

Ниже, на рисунке 14, представлен пример перевода шкалы пикселей в ангстремы для спектра из архива *СОПКА*. Рис. 14а иллюстрирует извлечённый с помощью программы Aladin из негатива, полученного с объективной призмой, спектр звезды HIP 88569 (в верхней части). Звезда имеет спектральный класс А0 и звёздную величину $8^m.34$ в полосе V. Сокращение CrAOB008-11 соответствует тому, что исходный негатив имеет номер 11 в коробке 8 архива спектральных наблюдений КрАО. В нижней части рисунка представлен фотометрический разрез изображения спектра.

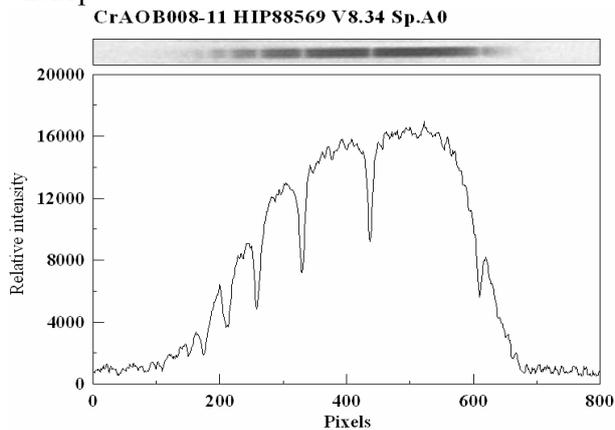
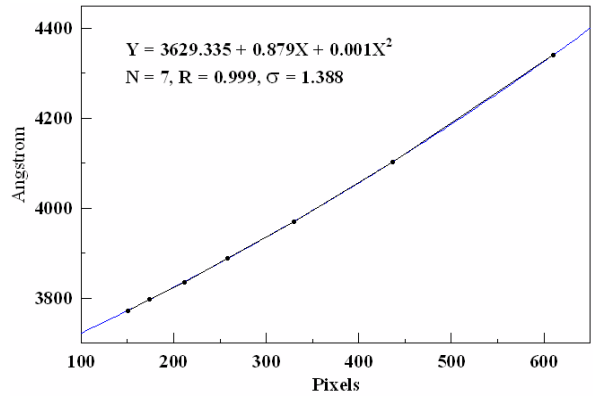

Рис. 14а.                                                                                          Рис. 14б.

На рис. 14б приведена калибровочная кривая для редукции шкалы пикселей в ангстремы, построенная по центрам тяжести основных спектральных линий HIP 88569 на уровне половины интенсивности в зависимости от соответствующей длины волны. Аппроксимирующая функция представляет собой полином второй степени с коэффициентами, найденными методом наименьших квадратов по семи значениям пиксел – длина волны. Коэффициент корреляции данных близок к единичному значению при ошибке в определении длины волны ~ 1 Å.

Рисунок 14в представляет собой финальный результат трансформации шкалы пикселей в изображении спектра низкой дисперсии в шкалу длин волн, выраженную в ангстремах.

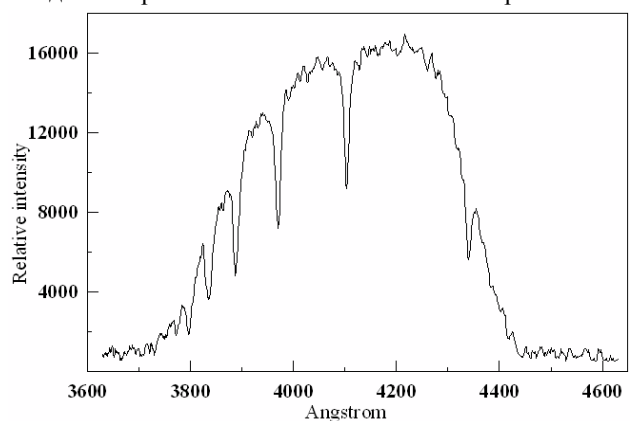

Рис. 14в.

## 3.3. Перевод шкалы пикселей в ангстремы для спектров с высокой дисперсией

Трансформация шкалы пикселей в ангстремы осуществлялась по спектру сравнения (полый железный катод с аргоновым наполнением), который впечатан над и под спектрограммой AG Peg (рис. 15а). Для построения дисперсионной кривой (пример на рис. 11) положение центров тяжести профилей избранных спектральных линий верхнего и нижнего фотометрического разреза спектра сравнения усреднялись с целью устранения возможного перекоса изображения. Результат перевода шкалы пикселей в ангстремы показан на рисунке 15б.

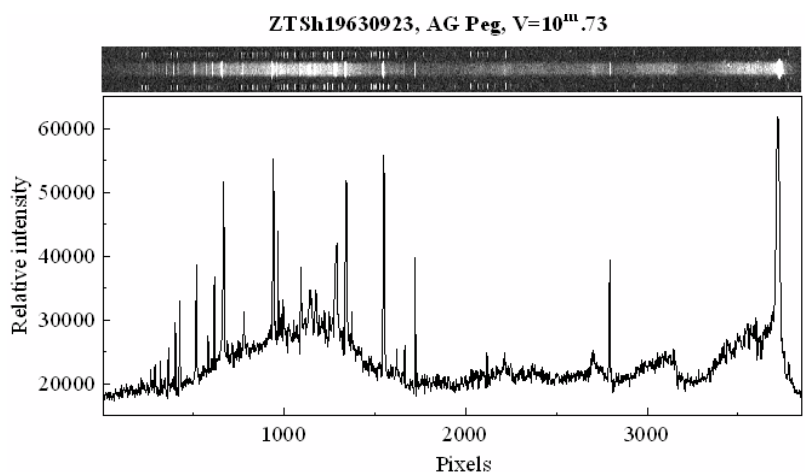

Рис. 15а.

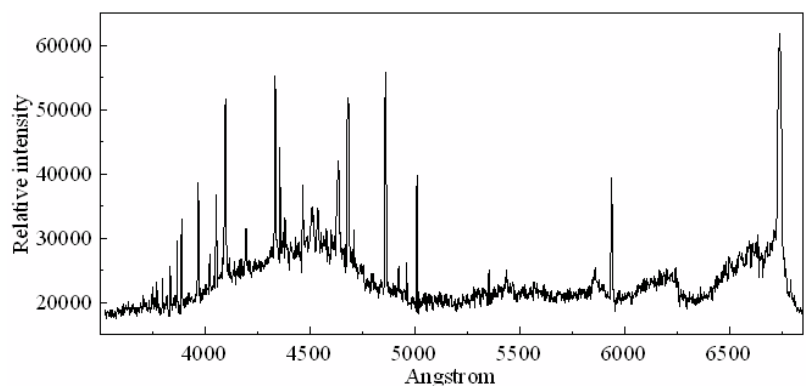

Рис. 15б.

## 4. Интерактивный доступ и обработка цифровых версий спектров
### 4.1. Представление информации в HTML формате

На рисунке 16 приведен фрагмент HTML страницы, обеспечивающей доступ к цифровой версии спектрограмм из архива академика А.А. Боярчука. Здесь показаны дата получения негатива и его номер по журналу наблюдений, изображения уменьшенных копий негативов, гиперссылки (обозначены на рис. 16 синим цветом) к отсканированной фотопластинке в JPG и FIT форматах, переход к данным фотометрического разреза и отображающему его рисунку.

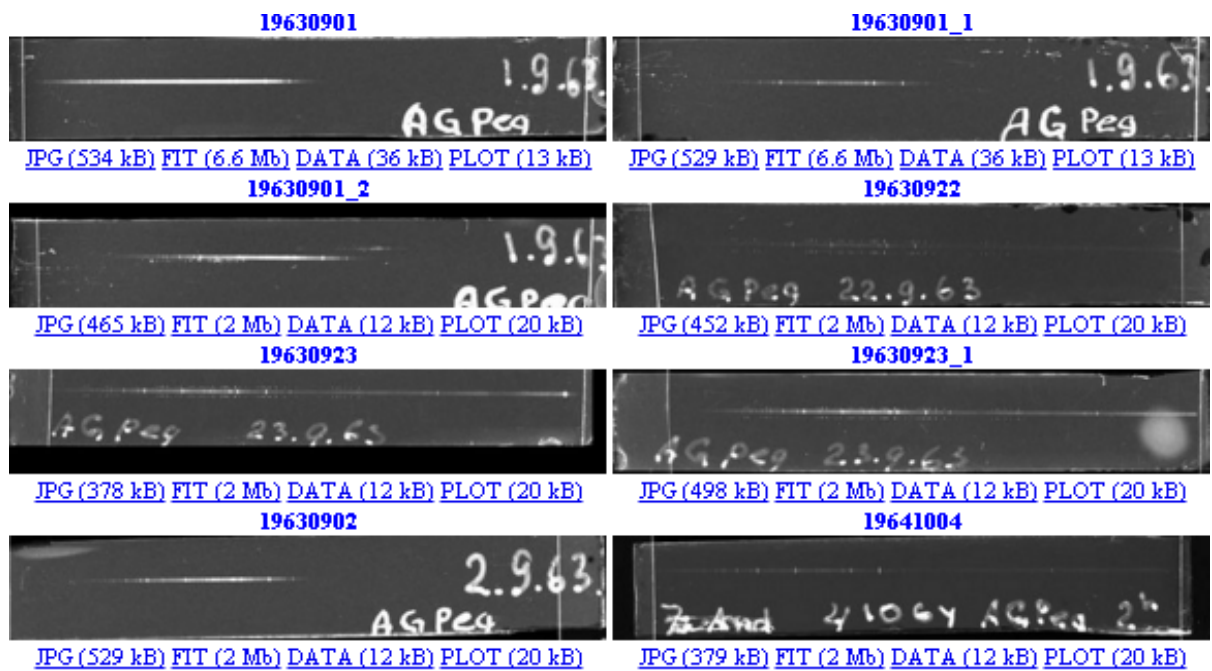

Рис. 16.

### 4.2. Идентификация спектральных линий

Для проведения идентификации спектральных линий в отсканированном спектре AG Peg, его фотометрический разрез был подготовлен в формате, поддерживаемом программой *Specview*, разработанной в Научном Институте Космического Телескопа (Буско, 2000). Программа является приложением Международной виртуальной обсерватории и обеспечивает интерактивный доступ к размещённой в сети информации, в том числе к спектральным архивам КрАО.

Из набора спектральных линий, содержащихся в базе данных *Specview*, отбирались те, которые наиболее хорошо видны. На рис. 17 показан интерфейс программы и указаны химические элементы с соответствующими им длинами волн. Синим цветом, показаны небулярные линии.

Отметим, что независимая идентификация спектральных линий по оцифрованному спектру AG Peg хорошо согласуется с информацией, опубликованной для этой звезды в 35 томе «Известий Крымской астрофизической обсерватории». В статье А.А. Боярчука и Р.Е. Гершберга «Спектроскопические наблюдения симбиотических звёзд Z And, AG Peg и AG Dra в 1963 г.» (Боярчук, Гершберг, 1966) приведена информация об относительных интенсивностях 50 спектральных линий AG Peg. На рис. 17 указаны 42 наиболее хорошо видимые спектральные линии на одном из оцифрованных архивных негативов.

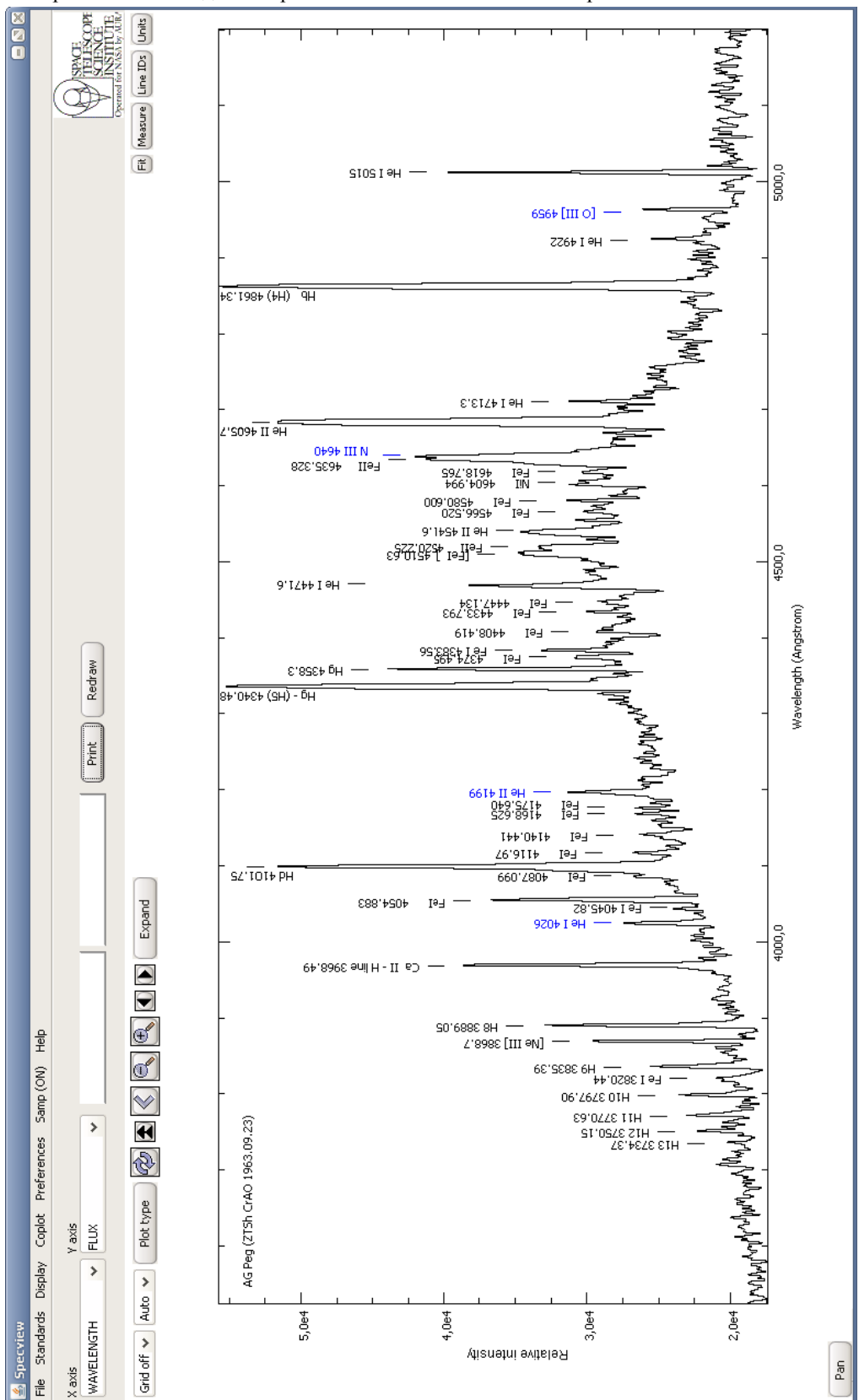

Рис. 17.

Больше примеров цифровых версий фотографических спектральных архивов описано нами в статье «Examples of Digital Versions of the CRAO Spectral Photographic Archives» (Горбунов, Шляпников, 2013).

## 4 Заключение

При выполнении работы, описанной в статье, продолжена каталогизация спектральных наблюдений КрАО. Отработана процедура сканирования негативов на сканере Epson Perfection V370 Photo, их калибровка, написание HTML, VOTable и AJS файлов для on-line доступа к архивной информации. На портале Крымской астрономической виртуальной обсерватории создана специальная страница, посвящённая цифровым спектральным архивам КрАО, доступ к которой можно получить по ссылке:

http://www.crao.ru/~aas/Spectral_Digital_Archives/CrAVO_SDA.html

## Литература

Боннарель и др. (Bonnarel F. et al.) // Astron. Astrophys. Suppl. Ser. 2000. V. 143. P. 33.
Боярчук А.А. // Изв. Крымск. Астрофиз. Обсерв. 2013. Т. 109. № 2. С. 113.
Борячук А.А., Гершберг Р.Е. // Изв. Крымск. Астрофиз. Обсерв. 1966. Т. 35. С. 3.
Буско (Busko I.) // ASP Conf. Ser. 2000. 216. 79.
Гершберг Р.Е. // Изв. Крымск. Астрофиз. Обсерв. 1995. Т. 90. С. 97-109.
Горбунов М., Шляпников А. (Gorbunov M., Shlyapnikov A.) // OAP. 2013. V. 26. P. 229.
Горбунов М.А., Шляпников А.А. // Изв. Крымск. Астрофиз. Обсерв. 2017а. Т. 113. № 1. С. 10.
Горбунов М.А., Шляпников А.А. // Изв. Крымск. Астрофиз. Обсерв. 2017б. Т. 113. № 1. С. 20.
Гранкин К.Н. // Изв. Крымск. Астрофиз. Обсерв. 2013. Т. 109. № 2. С. 105.
Долгов А.А., Шляпников А.А. // Изв. Крымск. Астрофиз. Обсерв. 2013. Т. 109. № 2. С.165.
Иоаннисиани Б.К., Тамбовский Г.А., Коншин В.М. // Изв. Крымск. Астрофиз. Обсерв. 1976. Т. LV. С. 208.
Копылов И.М. // Изв. Крымск. Астрофиз. Обсерв. 1954. Т. 11. С. 44.
Крючков С.В. и др. // Изв. Крымск. Астрофиз. Обсерв. 2009. Т. 104. № 6. С. 188.
Пакуляк Л. И др. (Pakuliak L., Shlyapnikov A., Rosenbush A. and Gorbunov M.) // ASInC. 2014. V. 11. P. 103.
Полосухина Н.С. и др. (Polosukhina N.S., Malanushenko V.P., Galkina T.S., Yavorskaya N.I.) // BCrAO. 1998. V. 94. P. 224.
Проник И.И. (Pronik I.I.) // KFNTS. 2005. V. 5. P. 250.
Рачковская Т.М. // Изв. Крымск. Астрофиз. Обсерв. 2013. Т. 109. № 2. С.118.
Шайн Г.А. // Изв. ГРАО. 1926. Т. 10. № 97. С. 450.
Шайн Г.А. // Изв. ГРАО. 1929. Т. 11. № 103. С. 199.
Шайн Г.А., Альбитский В.А.(Shajn G., Albitzky V.) // MNRAS. 1932. V. 92. P. 771.
Шляпников А.А. // Изв. Крымск. Астрофиз. Обсерв. 2013. Т. 109. № 2. С. 169.